\documentclass[aps,prl,twocolumn]{revtex4}

\usepackage{graphicx}% Include figure files
\usepackage{dcolumn}% Align table columns on decimal point
\usepackage{bm}% bold math
\usepackage{bezier}

\begin{document}

\title{Intrinsic dissipation in superconducting junctions probed by qubit spectroscopy}
\author{Dmitry S. Golubev$^{1}$  Artem V. Galaktionov$^{2}$,
and Andrei D. Zaikin$^{3,4}$
}
\affiliation{$^1$Low Temperature Laboratory, Department of Applied Physics, Aalto University, Espoo, Finland\\
$^2$I.E.Tamm Department of Theoretical Physics, P.N.Lebedev
Physical Institute, 119991 Moscow, Russia\\
$^3$Institute of Nanotechnology, Karlsruhe Institute of Technology (KIT), 76021 Karlsruhe, Germany\\
$^4$National Research University Higher School of Economics, 101000 Moscow, Russia}

\begin{abstract}
We propose to study frequency dependent intrinsic dissipation in a highly transparent Josephson junction 
by means of qubit spectroscopy. The spectral density of the effective dissipative bath may
contain significant contributions from Andreev bound states coupled to fluctuations of the Josephson phase.
Varying either the bias current applied to the junction or
magnetic flux through a superconducting ring in the rf-SQUID setup, one can tune the level splitting value
close to the bottom of the Josephson potential well. Monitoring the qubit
energy relaxation time one can probe the spectral density of the effective dissipative bath
and unambiguously identify the contribution emerging from Andreev levels.
\end{abstract}

%\pacs{}

\maketitle

\section{Introduction}

Highly transparent Josephson junctions have become an object of intense research in recent years.
Various types of such junctions have been studied: junctions based on carbon nanotubes \cite{Tinkham,CNK,Pillet},
aluminum atomic break junctions \cite{Saclay}, graphene \cite{Calado,Lee}, InAs nanowires \cite{Delsing,Marcus,Devoret}, 2d \cite{Molenkamp} and 3d \cite{Brinkman,Chalmers} topological insulators.
The physics of highly transparent Josephson junctions is determined by Andreev bound states, or Andreev levels,
which have the energies inside the supeconducting gap and take part in transferring the supercurrent.
The interest to this topic is mostly triggered by possible applications in quantum computing. One of the proposals, for example, suggests to use two Andreev levels as qubit states \cite{Shumeiko}. More elaborate proposals involve Majorana states \cite{Kitaev} formed in junctions with 2d and 3d topological insulators and InAs nanowires in presence of the magnetic field. 
Coherent manipulation of the populations of Andreev bound states has been demonstrated
in recent experiments with aluminum break junctions \cite{Saclay2} and InAs nanowires \cite{Devoret}.  

In this letter we propose to use the usual setup, in which a qubit is coupled to a readout resonator \cite{Koch}, to perform
spectroscopy of Andreev levels in a highly transparent Josephson junction.
This technique has become standard to measure the environmental noise spectrum affecting a qubit,
cf., e.g., \cite{Martinis},
and it has also been employed to study random switching between Andreev levels
\cite{Devoret}. We have recently demonstrated \cite{PRB} that Andreev levels couple to
fluctuations of the Josephson phase and, hence, introduce extra damping in our system.
In other words, such low energy Andreev states can form an {\it intrinsic} effective quantum dissipative environment for the Josephson phase $\varphi$
which can strongly modify quantum dynamics of superconducting weak links as
compared to that for tunnel junctions. For instance, macroscopic quantum tunneling of the Josephson phase $\varphi$ acquires a number of novel 
features \cite{PRB,We2,we3}.
Note that  similar situation also occurs, e.g., in junctions made of $d$-wave superconductors, where the so-called midgap 
Andreev bound states are formed 
\cite{Shumeiko1}, and even in low transparency Josephson junctions, thus influencing the relaxation rate of usual qubits \cite{Catelani}. 
Here we are going to study the effect of this intrinsic damping on the energy relaxation time (the so-called $T_1$-time) of a qubit
containing highly transparent Josephson junction.
We will propose to use a qubit setup as a spectroscopic tool in order to resolve
frequency dependent dissipation in the junction.

Below we will analyze the behavior of three different types of superconducting weak links:
a junction with few highly transparent channels, a short superconductor - insulator - normal metal - insulator - superconductor junction (SINIS junction),
and a short superconductor - normal metal - superconductor junction (SNS junction).
For these three cases we derive the dependence of $T_1$ on the bias current applied to the junction.
Our results for $T_1$-time are also valid in the classical regime in which level quantization may be ignored.
In this case $T_1$ may be determined with the aid of routine microwave reflection measurements.

\section{Model and Theory}

\begin{figure}[!ht]
\includegraphics[width=\columnwidth]{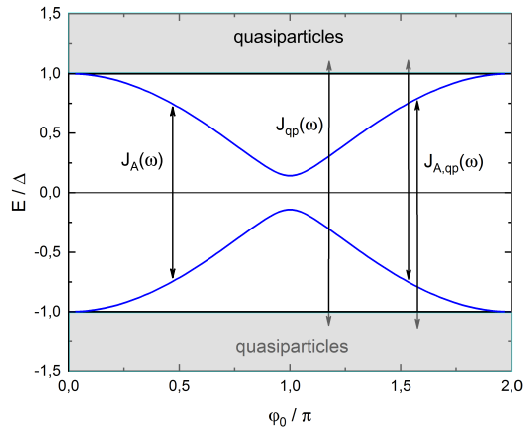}
\caption{Subgap Andreev levels (blue lines) inside a superconducting junction.  
Possible transitions at zero temperature between such levels and continuous quasiparticle spectrum are illustrated by vertical lines.
Arrows indicate the transitions contributing to the spectral densities $J_A(\omega)$, $J_{A,qp}(\omega)$ and $J_{qp}(\omega)$.}
\label{levels}
\end{figure}

The current-phase relation of a short Josephson junction characterized by an arbitrary distribution
of transmissions on the normal state conducting channels $\tau_n$ has the form \cite{KO}
\begin{eqnarray}
I_{J}(\varphi )=\frac{e\Delta^2\sin\varphi}{2}\sum_n\frac{\tau_n}{\epsilon_n(\varphi)}\tanh\frac{\epsilon_n(\varphi)}{2T},
\label{IJ}
\end{eqnarray}
where $\varphi$ is the Josephson phase, and
\begin{equation}
\epsilon_n(\varphi)=\Delta\sqrt{1-\tau_n\sin^2(\varphi/2)}.
\label{And}
\end{equation}
The energies of Andreev bound states inside the energy gap equal to $\pm\epsilon_n(\varphi)$, they are shown in Fig. \ref{levels}.

\begin{figure}[!ht]
\includegraphics[width=\columnwidth]{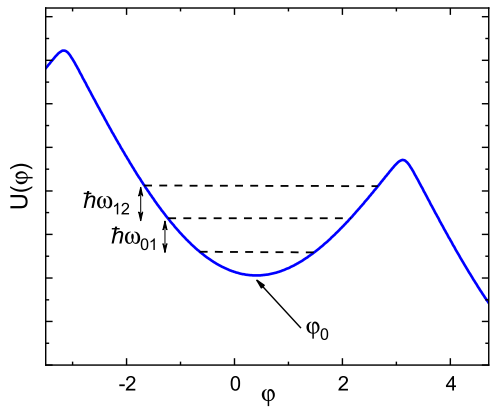}
\caption{Tilted Josephson potential and several energy levels at the bottom of the potential well.
The potential minimum is reached at the phase value $\varphi=\varphi_0$. The frequency $\omega_{01}$  ($\omega_{12}$) controls splitting between the levels 0 and 1 (1 and 2).}
\label{potential}
\end{figure}

The critical current is found by maximizing the expression (\ref{IJ}), $I_C=\max_\varphi\{I_J(\varphi)\}$.
At bias currents below $I_C$ phase dynamics is determined by tilted Josephson potential having the form
\begin{eqnarray}
U_J(\varphi)=-2T\sum_n \ln\left(\cosh\frac{\epsilon_n(\varphi)}{2T}\right) - \frac{I\varphi}{2e},
\label{UJ}
\end{eqnarray}
where $I$ is the bias current. The characteristic energy scale of the potential barrier is given by Josephson energy $E_J=I_C/2e$.
The potential (\ref{UJ}) is depicted in Fig. \ref{potential}.

If the junction has a capacitance $C$, its quantum Hamiltonian acquires the form \cite{SZ}
\begin{eqnarray}
\hat H_J = -4E_C\frac{\partial^2}{\partial\varphi^2} + U_J(\varphi),
\label{HJ}
\end{eqnarray}
where $E_C=e^2/2C$ is the charging energy. The value $C$ is defined as a sum
of geometric capacitance of both the junction and superconducting leads and the
renormalization term emerging from the coupling to subgap Andreev levels and
overgap quasiparticles \cite{we}.
Solving the Schr\"odinger equation with the Hamiltonian (\ref{HJ}) one can identify the
energy levels close to the bottom of the potential well.
The two lowest levels form a qubit. In the semiclassical limit $E_J\gtrsim E_C$,
the corresponding level splitting, $\omega_{01}$,
can be approximately evaluated as
\begin{eqnarray}
\omega_{01} = \sqrt{\frac{4E_C}{e}\frac{dI_J(\varphi_0)}{d\varphi}},
\label{omega01}
\end{eqnarray}
where $\varphi_0$ is the phase at which the potential has the minimum, see Fig. \ref{potential}.
It depends on the bias current and should be found from the equation
\begin{eqnarray}
I_J(\varphi_0)=I.
\end{eqnarray}
Eq. (\ref{omega01}) does not apply for bias currents very close to $I_C$. Namely, in the range of bias currents
\begin{eqnarray}
1-\frac{I}{I_C} \lesssim \frac{1}{2}\left(\frac{81}{16}\frac{E_C}{E_J}\right)^{2/5}
\label{cond1}
\end{eqnarray} 
only one localized state remains at the bottom of the potential well and qubit spectroscopy becomes impossible.

In order to quantify the anharmonicity of the potential we define the parameter
\begin{eqnarray}
\alpha = \omega_{12}-\omega_{01},
\end{eqnarray}
where $\omega_{12}$ defines splitting between the levels  1 and 2, see Fig. 2.
This anharmonicity parameter should be sufficiently large in order to decouple the transition between these
levels from both readout and control pulses at qubit frequency $\omega_{01}$.
At zero bias current the potential (\ref{UJ}) is quartic close to the bottom of the potential well,
and the anharmonicity parameter acquires the form
\begin{eqnarray}
\alpha = \frac{I'''_J(0)}{I'_J(0)}E_C.
\end{eqnarray}
At high bias the potential is characterized by cubic anharmonicty close to the bottom, in which case
\begin{eqnarray}
\alpha =  -\frac{5}{3}\left(\frac{I''_J(\varphi_0)}{I'_J(\varphi_0)}\right)^2E_C.
\end{eqnarray}

As we already pointed out, subgap Andreev levels form an effective environment for the Josephson phase $\varphi$ \cite{PRB}. In order to properly account for quantum dynamics of the system "particle + environment"
it is instructive to employ the Feynman-Vernon influence functional technique \cite{FH}. This quantum environment can be described phenomenologically in terms of harmonic oscillators \cite{FH,CL,Weiss} or microscopically \cite{AES,SZ} as electron sea in a disordered metal. Tracing out the environment degrees of freedom 
and assuming that they are at equilibrium characterized by temperature $T$ 
one arrives at the effective action describing quantum dissipation. More sophisticated versions of the influence functional \cite{GZ98,GZS} also account for Fermi statistics for electrons in a metal forming an effective environment "for themselves".

Recently we demonstrated \cite{PRB} that at low enough energies the effective action for the coupled system
"Josephson phase + Andreev levels" is equivalent to the effective Hamiltonian
\begin{eqnarray}
&& \hat H = \hat H_J
\nonumber\\ &&
+\,\sum_n \left[\frac{\hat
P_n^2}{2M_n}+\frac{M_n\omega_n^2}{2}\left(Q_n-\frac{c_n(\varphi-\varphi_0)}{M_n\omega_n^2}\right)^2\right],
\label{HQintphi}
\end{eqnarray}
expressed in the standard form describing a quantum particle $\varphi$ interacting with a collection of harmonic oscillators $Q_n$.
As we have pointed out above, the phase $\varphi_0$, appearing in the Hamiltonian, can be changed by the bias current.
This design corresponds to a phase qubit. 
Alternatively, one can change the phase $\varphi_0$ by embedding the junction in the superconducting ring
and changing the magnetic flux $\Phi$ through the ring, in which case $\varphi_0=2\pi\Phi/\Phi_0$ (here $\Phi_0$ is the flux quantum).
This rf-SQUID geometry corresponds to a flux qubit.

The Hamiltonian (\ref{HQintphi}) applies as long as the phase fluctuations remain weak, $\sqrt{\langle (\varphi-\varphi_0)^2\rangle}\ll 1$.
The second line of Eq. (\ref{HQintphi}) describes an effective Andreev level
environment which consists of harmonic oscillators with frequencies $\omega_n=2\epsilon_n(\varphi_0)$, corresponding to the difference between the energies of
the two Andreev levels $\pm\epsilon_n(\varphi_0)$
coupled to the "particle coordinate" $\varphi$. The coupling constants $c_n$ are given by the expression \cite{PRB,We2}
\begin{equation}
\frac{c_n^2}{M_n}=\tau_n^2(1-\tau_n) \frac{\Delta^4}{\epsilon_n(\varphi_0)}\sin^4\frac{\varphi_0}{2}\tanh\frac{\epsilon_n(\varphi_0)}{2T}.
\label{randg}
\end{equation}
The coupling constants $c_n$ depend on channel transmissions $\tau_n$, temperature, and the bias current via the equilibrium phase $\varphi_0$.
One finds that, e.g., at $T=0$ and for bias current values close to $I_C$ the maximum coupling for a single channel junction is achieved at the transmission value $\tau_n=8/9$.

Following \cite{CL,Weiss} let us define the effective environment frequency spectrum as
\begin{eqnarray}
J_A(\omega,\varphi_0) = \frac{\pi}{2}\sum_n\frac{c_n^2}{M_n\omega_n}\delta(\omega-\omega_n), \quad \omega_n=2\epsilon_n(\varphi_0).
\label{Jo}
\end{eqnarray}
In the case of a dissipative environment formed by Andreev levels this spectrum depends on the particular distribution of normal transmissions $\tau_n$ in the junction. At $T=0$ from Eqs. (\ref{randg}), (\ref{Jo})  we obtain
\begin{eqnarray}
J_{A}(\omega)=\sum_n \frac{\pi\Delta^4 \tau_n^2(1-\tau_n)\sin^4\frac{\varphi_0}{2}}{4\epsilon_n^2(\varphi_0)}\delta(\omega-2\epsilon_n(\varphi_0)).
\label{JA0}
\end{eqnarray}

Up to now we have considered an ideal situation assuming that Andreev levels have an infinite lifetime.
In practice, however, their lifetime remains finite due to dephasing processes. Here we introduce
the dephasing rate, $\gamma_\varphi$, on phenomenological level replacing the $\delta-$functions
in the spectral density (\ref{JA0}) by Lorentzians with an effective width $2\gamma_\varphi$. Then the
spectral density of our effective environment becomes
\begin{eqnarray}
J_{A}(\omega)=\sum_n \frac{\gamma_\varphi\Delta^4 \tau_n^2(1-\tau_n)\sin^4\frac{\varphi_0}{2}}
{2\epsilon_n^2(\varphi_0)\left[ (\omega-2\epsilon_n(\varphi_0))^2 + 4\gamma_\varphi^2 \right]}.
\label{JA}
\end{eqnarray}
This expression should be employed in the range of parameters where the spectral
density (\ref{JA0}) equals to zero.

Andreev levels give the dominating contribution to the junction dissipation at subgap frequencies. At higher frequencies two extra mechanisms may also contribute \cite{we,we2}.
The first of these mechanisms is related to possible transitions between Andreev levels and continuous quasiparticle spectrum at energies above $\Delta$ (see Fig. \ref{levels}). At zero temperature the corresponding spectral density reads \cite{we}
\begin{eqnarray}
J_{A,qp}(\omega) &=& \sum_n  \frac{\tau_n^{3/2}}{4}
\theta(\omega-\Delta-\epsilon_n(\varphi_0))
\nonumber\\ &&\times\,
\frac{\Delta\sin\frac{\varphi_0}{2} (\omega\epsilon_n(\varphi_0)-\Delta^2(1+\cos\varphi_0))}{\epsilon_n(\varphi_0)[(\omega-\epsilon_n(\varphi_0))^2-\epsilon_n^2(\varphi_0)]}
\nonumber\\ &&\times\,
\sqrt{(\omega-\epsilon_n(\varphi_0))^2-\Delta^2}.
\label{Aqp}
\end{eqnarray}
The second mechanism is associated with dissipation by quasiparticles (again see Fig. \ref{levels}).
At $T=0$ the quasiparticle spectral density has the form \cite{we}
\begin{eqnarray}
J_{qp}(\omega) &=& \theta(\omega-2\Delta)\sum_n \frac{\tau_n}{2}\int_{\Delta}^{\omega-\Delta}\frac{dE}{2\pi}
\nonumber\\ &&\times\,
\left[ E(\omega-E) - \Delta^2\cos\varphi_0 - \tau_n\Delta^2\sin^2\frac{\varphi_0}{2} \right]
\nonumber\\ && \times\,
\frac{\sqrt{E^2-\Delta^2}\sqrt{(\omega-E)^2-\Delta^2}}{(E^2-\epsilon_n^2(\varphi_0))[(\omega-E)^2-\epsilon_n^2(\varphi_0)]}.
\label{qp}
\end{eqnarray}
Hence, the total spectral density of the effective oscillator bath, giving rise to intrinsic dissipation in the junction, reads
\begin{eqnarray}
J(\omega)=J_A(\omega)+J_{A,qp}(\omega)+J_{qp}(\omega).
\end{eqnarray}

The effect of the environment on the dynamics of the Josephson phase can also be mimicked by introducing a linear
shunting impedance $Z_S(\omega)$ connected in parallel to the junction.
The spectral density $J(\omega)$ can be related to this impedance by means of the formula
\begin{eqnarray}
{\rm Re}\left(\frac{1}{Z_S(\omega)}\right) = 4e^2\frac{J(\omega)}{\omega}.
\end{eqnarray}
The properties of the impedance $Z_S(\omega)$ have been studied in detail in Ref. \cite{Glazman}. 

Finally, let us introduce the energy relaxation time of the junction $T_1$. It is defined as
\begin{eqnarray}
\frac{1}{T_1}=\frac{8E_C }{\omega_{01}}J\big(\omega_{01},\varphi_0\big)\coth\frac{\omega_{01}}{2T}.
\label{gammain}
\end{eqnarray}
This $T_1$-time describes both classical dissipative dynamics of the Josephson phase
and relaxation between the two states of the qubit. In the normal state with $\Delta\to 0$ Eq. (\ref{gammain}) yields
the expected result $T_1=R_NC$, where $R_N$ is the normal state junction resistance.

\section{Examples}

\subsection{Nanowire with few conducting channels}

An important exactly solvable example corresponds to a junction made of two superconducting leads
connected by a short nanowire with few conducting channels.
We assume that all these channels are characterized the same transmission value $\tau$.
Then the zero temperature critical current takes the form
\begin{eqnarray}
I_C = \frac{2\pi{\cal N}\tau}{1+\sqrt{1-\tau}} \frac{\Delta}{eR_q},
\end{eqnarray}
where ${\cal N}$ is the total number of conducting channels, and $R_q=h/e^2$ is the resistance quantum.
The corresponding Josephson energy is
\begin{eqnarray}
E_J=\frac{I_C}{2e}=\frac{{\cal N}\tau}{1+\sqrt{1-\tau}} \frac{\Delta}{2}.
\end{eqnarray}
The critical current is achieved at the following value of the phase:
\begin{eqnarray}
\varphi_0^{\rm cr}=\pi-\arccos\left[\frac{\left(1-\sqrt{1-\tau}\right)^2}{\tau}\right].
\end{eqnarray}
\begin{figure}[!ht]
\includegraphics[width=\columnwidth]{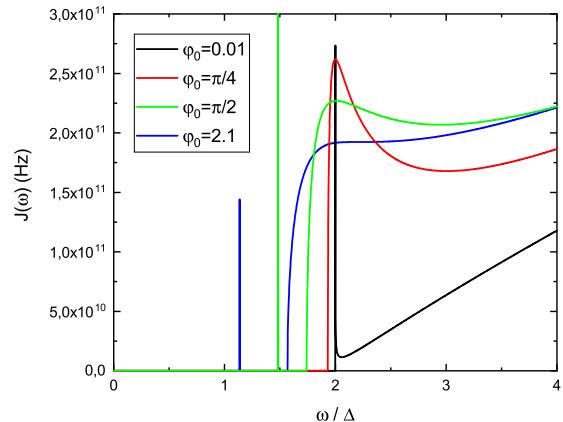}
\caption{The effective bath spectral density for a junction with two conducting channels of transmission $\tau=0.9$ for four different values of the phase $\varphi_0$. The
linewidth of Andreev levels is supposed to be independent on the bias current and equals $\gamma_\varphi=200$ kHz, the superconducting gap is $\Delta=200$ $\mu$eV.
Sharp peak is due to a doubly degenerate Andreev level.}
\label{JCN}
\end{figure}

\begin{figure}[!ht]
\includegraphics[width=\columnwidth]{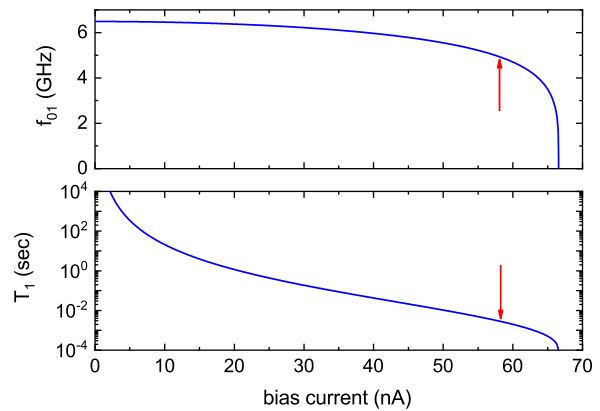}
\caption{The frequency $f_{01}=\omega_{01}/2\pi$ (top panel) and the $T_1$-time induced by coupling to Andreev levels (bottom panel)
versus bias current for a nanowire with two conducting channels. Charging energy is set to be $E_C=1$ $\mu$eV, other parameters are the same as
in Fig. \ref{JCN}. 
Red arrows in both plots indicate the value of the bias current (\ref{cond1}) above which less than two levels exists in the potential well,
and qubit spectroscopy becomes impossible.
}
\label{T1_CN}
\end{figure}

Qubit level splitting in this regime is estimated as
\begin{eqnarray}
\omega_{01}=\left\{
\begin{array}{ll}
2\sqrt{1+\sqrt{1-\tau}}\sqrt{E_JE_C}, & I=0,
\\
2^{7/4}\sqrt{E_JE_C}\left(1-I/I_C\right)^{1/4}, & I\to I_C,
\end{array}
\right.
\end{eqnarray}
and the anharmonicity parameter is found to be
\begin{eqnarray}
\alpha=\left\{
\begin{array}{ll}
-\left(1-\frac{3}{4}\tau \right)E_C, & I=0,
\\
-\frac{5E_C}{6(1-I/I_C)}, & I\to I_C.
\end{array}
\right.
\end{eqnarray}
Provided the junction is operated as a qubit, its frequency $\omega_{01}/2\pi$ is usually tuned to the range from $2$ to $5$ GHz.
This value is significantly lower than the frequency corresponding to the gap of aluminum ($\Delta\approx 200$ $\mu$eV = 48 GHz).
As a consequence, the inequality $\omega_{01}<2\Delta\sqrt{1-\tau\sin^2(\varphi_0/2)}$ holds at all bias current values $0<I<I_C$.
In this case the spectral density of the smeared Andreev levels (\ref{JA}) gives the main contribution to
the relaxation time $T_1$ (\ref{gammain}).  To be specific, we consider a Josephson junction
with aluminum leads connected by a short single wall carbon nanotube or InAs nanowire with two conducting channels having transmission probability $\tau=0.9$
and assume that the charging energy has the value $E_C=1\;\mu{\rm eV}=2\pi\times 240$ MHz, which is typical for qubits shunted
by a large capacitance.
For these parameters we find the amplitude of zero point phase fluctuations equal to
$$
\sqrt{\langle(\varphi-\varphi_0)^2\rangle}=(2E_C/E_J)^{1/4}=0.348 <1,
$$
thus assuring that the simple Hamiltonian (\ref{HQintphi}) remains applicable with a sufficient accuracy.
The dephasing rate $\gamma_\varphi$ depends on both temperature and the phase $\varphi_0$.
Detailed analysis of this dependence is beyond the scope of this paper.
For illustrative purpose here we assume that $\gamma_\varphi$  roughly equals to the switching rate between two Andreev levels
observed in the experiment with an InAs nanowire \cite{Devoret}, $\gamma_\varphi/2\pi \approx 200 $ kHz.
The results of this approximation are plotted in Fig. \ref{T1_CN}, where we show both the frequency $f_{01}=\omega_{01}/2\pi$
and the time $T_1$ at zero temperature. Coupling to Andreev levels vanishes at zero bias, that is why
$T_1$ is diverging at this point. On the contrary, at bias currents close to $I_C$ the time $T_1$ becomes shorter.
Overall, our estimate gives rather long $T_1$-times. Hence, in experiment other relaxation mechanisms would dominate
unless special care is taken.

\subsection{SINIS junction}

\begin{figure}[!ht]
\includegraphics[width=\columnwidth]{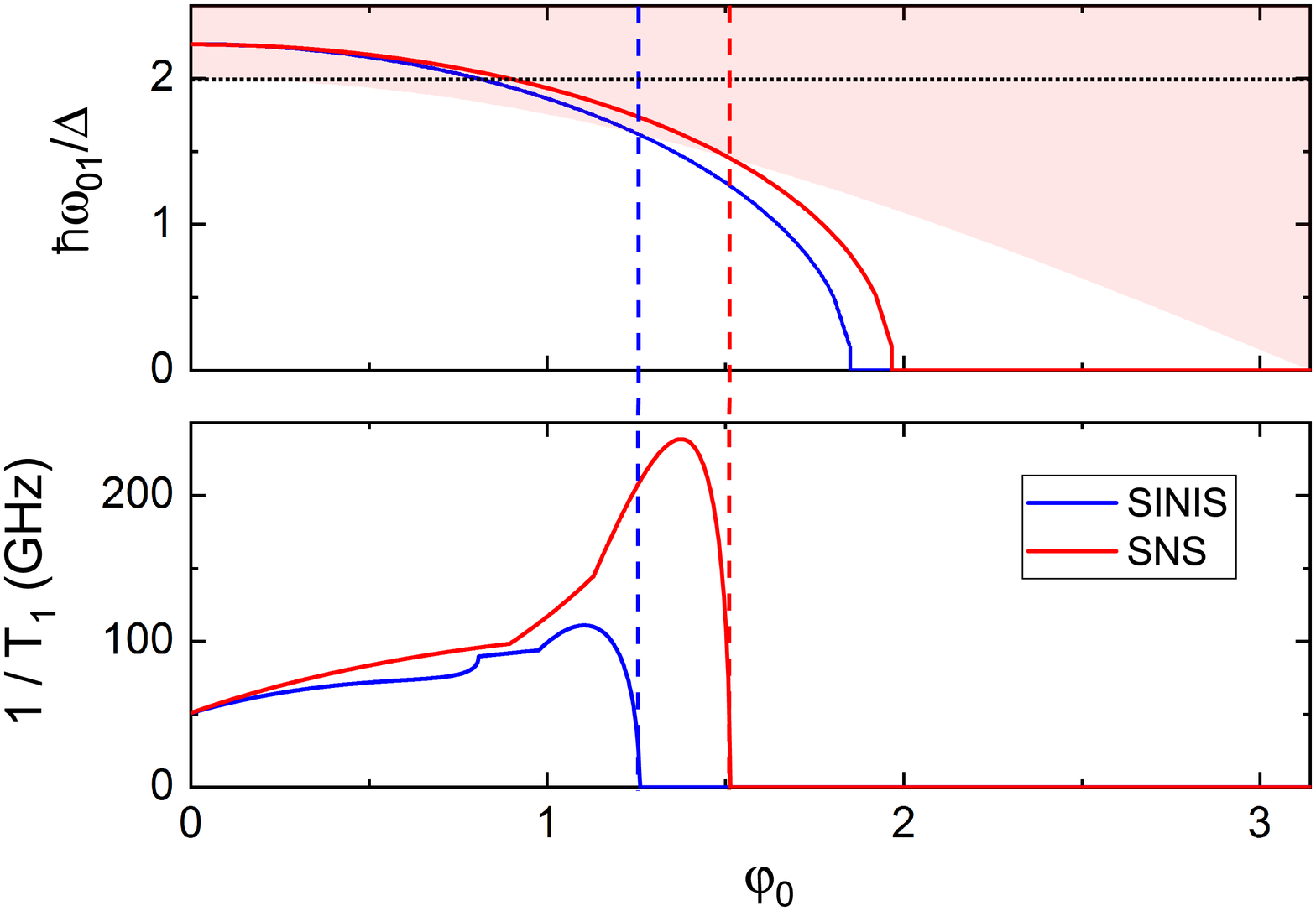}
\caption{The frequency $\omega_{01}$ (upper panel) and the energy relaxation rate $1/T_1$ (lower panel) as  functions of the
phase $\varphi_0$ for an SINIS and SNS junctions at $T=0$. Both junctions have the same parameters:  $\Delta=200$ $\mu$eV, $E_C=10$ $\mu$eV, $R_n=260$ $\Omega$.
Shaded area in the upper panel indicates the region $\omega_{01}>2\Delta\cos(\varphi_0/2)$, where intrinsic dissipation is expected to be strong.}
\label{T1}
\end{figure}

Another important exactly solvable limit corresponds to a short SINIS junction
with many conducting channels.
It is characterized by the following distribution of transmission eigenvalues \cite{Shep}:
\begin{eqnarray}
P(\tau)=\sum_n\delta(\tau-\tau_n)=\frac{R_q}{2\pi R_N}\frac{1}{\sqrt{\tau^3(1-\tau)}},
\end{eqnarray}
where $R_N$ is the normal state resistance given by Landauer formula,
$1/R_N=(2/R_q)\sum_n\tau_n$.
At zero temperature the Josephson current acquires the form
\begin{eqnarray}
I_J(\varphi)=\frac{\Delta}{eR_N}K\left(\left|\sin\frac{\varphi}{2}\right|\right)\sin\varphi,
\end{eqnarray}
where $K(x)$ is the complete elliptic integral of the first kind.
The critical current value equals to $I_C\approx 1.915 \Delta/eR_N$ and is achieved at the phase value $\varphi_0^{\rm cr}=1.854$.
The qubit frequency in this case reads
\begin{eqnarray}
\omega_{01}=\left\{ \begin{array}{cc}
\sqrt{\frac{E_C\Delta}{R_qR_N}} = 2.561\sqrt{E_JE_C}, & I=0,
\nonumber\\
3.215\sqrt{E_JE_C}\left(1-\frac{I}{I_C}\right)^{1/4}, & I\to I_C.
\end{array}\right.
\end{eqnarray}
This frequency is plotted in the top panel of Fig. \ref{T1}.

\begin{figure}[!ht]
\includegraphics[width=\columnwidth]{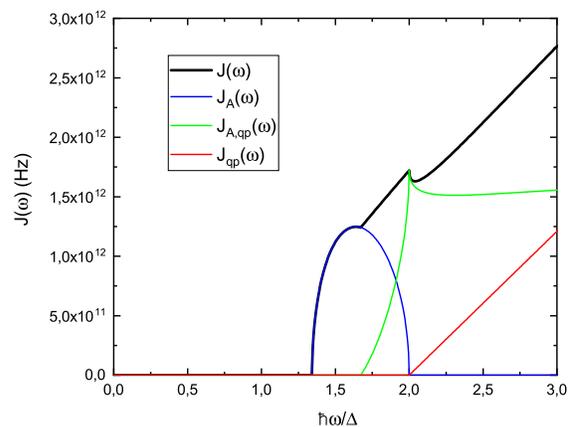}
\caption{The effective bath spectral density for an SINIS junction at $T=0$.
Various contributions are illustrated by different colors.
The junction parameters of the junction are chosen as follows: $\Delta=200$ $\mu$eV, $E_C=10$ $\mu$eV, $R_n=260$ $\Omega$ and $I_C=1.7$ $\mu$A. The phase value is set to be $\varphi_0=1.67$. We have used the approximation (\ref{Jqp_app}) for the quasiparticle spectral density.}
\label{J}
\end{figure}

\begin{figure}[!ht]
\includegraphics[width=\columnwidth]{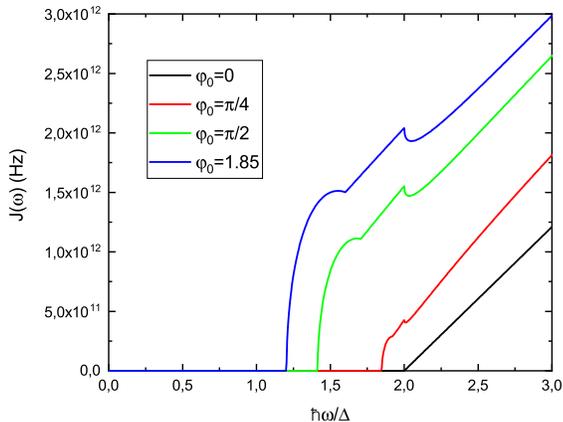}
\caption{The same as in Fig. \ref{J} for various phase values $\varphi_0$.}
\label{Jphi}
\end{figure}

The anharmonicity parameter is
\begin{eqnarray}
\alpha=\left\{
\begin{array}{ll}
-\frac{5}{8}E_C, & I=0,
\\
-0.696\frac{E_C}{1-I/I_C}, & I\to I_C.
\end{array}
\right.
\end{eqnarray}

The spectral density of Andreev levels takes the form
\begin{eqnarray}
J_A(\omega)=\frac{R_q}{16R_N}\theta\left(\omega-2\Delta\cos\frac{\varphi_0}{2}\right)\theta(2\Delta-\omega)
\nonumber\\
\frac{\sqrt{4\Delta^2-\omega^2} \sqrt{\omega^2-4\Delta^2\cos^2\frac{\varphi_0}{2}} }{\omega}.
\end{eqnarray}
The quasiparticle spectral density (\ref{qp}) can be approximate as
\begin{eqnarray}
J_{qp}(\omega) \approx \frac{R_q\theta(\omega-2\Delta)(\omega-2\Delta) }{8\pi R_N}.
\label{Jqp_app}
\end{eqnarray}
For simplicity, we use this approximation in the rest of this paper. 
The total spectral density of an SINIS junction is plotted in Figs. \ref{J} and \ref{Jphi}.
It equals to zero below the threshold frequency
\begin{eqnarray}
\omega_{\rm th}=2\Delta\cos\frac{\varphi_0}{2},
\label{omega_th}
\end{eqnarray}
which depends on the phase $\varphi_0$ and, hence, on the bias current.

Provided both $E_J$ and $E_C$ are sufficiently large,
the qubit frequency $\omega_{01}$ can cross the threshold (\ref{omega_th})
at certain value of the bias current. This situation is illustrated in the upper panel of Fig. \ref{T1}.
Should that happen, the $T_1$-time becomes very short, and the quality factor of the junction gets reduced a lot
since the dissipation rate $1/T_1$ becomes comparable to the frequency $\omega_{01}/2\pi$, 
see the bottom panel of Fig. \ref{T1} as well as Fig. \ref{T1I}.
This effect should be easily detectable in microwave reflection measurements.

\subsection{SNS junction}

\begin{figure}[!ht]
\includegraphics[width=\columnwidth]{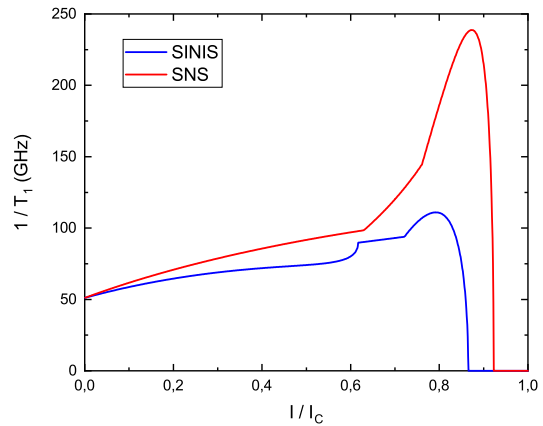}
\caption{Inelastic relaxation rate $1/T_1$ as a function of the bias current. The junction parameters are the same as in Fig. \ref{T1}.}
\label{T1I}
\end{figure}

Let us now briefly consider yet another important junction type, a short superconductor - normal metal - superconductor (SNS) junction.
In this case the distribution of channel transmissions reads \cite{Dorokhov}
\begin{eqnarray}
P(\tau)=\sum_n\delta(\tau-\tau_n)=\frac{R_q}{4 R_N}\frac{1}{\tau\sqrt{1-\tau}},
\end{eqnarray}
and the current phase relation at $T=0$ takes the form \cite{Kulik}
\begin{eqnarray}
I_J(\varphi)=\frac{\pi\Delta}{eR_N}\cos\frac{\varphi}{2}\,{\rm arctanh}\left(\sin\frac{\varphi}{2}\right).
\end{eqnarray}
For this junction the critical current equals to $I_C\approx 2.082 \Delta/eR_N$ and is achieved at the phase value $\varphi_0^{\rm cr}=1.971$. The frequency $\omega_{01}$ is approximated as follows
\begin{eqnarray}
\omega_{01}=\left\{ \begin{array}{cc}
\sqrt{\frac{E_C\Delta}{R_qR_N}} = 2.457\sqrt{E_JE_C}, & I=0,
\nonumber\\
3.2\sqrt{E_JE_C}\left(1-\frac{I}{I_C}\right)^{1/4}, & I\to I_C.
\end{array}\right.
\end{eqnarray}
Evaluating the anharmonicity parameter we obtain
\begin{eqnarray}
\alpha=\left\{
\begin{array}{ll}
-E_C/2, & I=0,
\\
-0.683\frac{E_C}{1-I/I_C}, & I\to I_C,
\end{array}
\right.
\end{eqnarray}
and the spectral density of Andreev levels reads
\begin{eqnarray}
J_A(\omega)=\frac{\pi R_q}{64R_N}\theta\left(\omega-2\Delta\cos\frac{\varphi_0}{2}\right)\theta(2\Delta-\omega)
\nonumber\\
\frac{\left(4\Delta^2-\omega^2\right) \sqrt{\omega^2-4\Delta^2\cos^2\frac{\varphi_0}{2}} }{\Delta\omega\sin\frac{\varphi_0}{2}}.
\end{eqnarray}

Similarly to the case of an SINIS junction, here the frequency $\omega_{01}$ may also cross the threshold (\ref{omega_th}) provided the junction is not artificially shunted
by a capacitor and its charging energy is substantial. In this case the energy relaxation time become very short as it is illustrated in Figs. \ref{T1} and \ref{T1I}.

We also note that icrowave response of long SNS junctions with Thouless energies lower than $\Delta$ has been 
studied in Refs. \cite{Virtanen,Feigelman}.

\section{Summary}

We have analyzed the parameters of phase or flux qubits based on a highly transparent Josephson junction.
We have found that for a junction with few conducting channels and with low charging energy dissipation
caused by coupling of the fluctuating phase to subgap Andreev levels is rather weak.
In this case it is determined by a finite width of Andreev levels induced by dephasing processes. We have also considered short SINIS and SNS junctions.
In this case the qubit frequency $\omega_{01}$ may cross the low frequency dissipation threshold (\ref{omega_th})
at certain value of the bias current, which should easily be detectable. We believe that qubit
spectroscopy can serve as a useful tool to reveal the coupling between Andreev levels and the Josephson phase.

This work was supported in part by RFBR Grant No. 18-02-00586.

\end{document}